\documentclass[lettersize,journal]{IEEEtran}
\usepackage{amsmath,amsfonts}
\usepackage{algorithmic}
\usepackage{algorithm}
\usepackage{array}
\usepackage{textcomp}
\usepackage{stfloats}
\usepackage{url}
\usepackage{verbatim}
\usepackage{graphicx}
\usepackage{cite}
\usepackage{amsmath}
\usepackage{color,multicol}
\usepackage{mathtools} 
\usepackage{amsfonts} 
\usepackage{array}
\usepackage{amssymb}
\usepackage{float}
\usepackage{epstopdf}
\usepackage{siunitx}
\usepackage[utf8]{inputenc}
\usepackage[english]{babel}
\usepackage{lipsum}
\usepackage{fancyhdr}
\usepackage{subcaption}

\usepackage{algorithmic}

\usepackage{algorithm}

\begin{document}

\title{Relative Entropy (RE) Based LTI System Modeling Equipped with time delay Estimation and Online Modeling}

\author{Mahdi Shamsi, Soosan Beheshti
\thanks{Mahdi Shamsi and Soosan Beheshti are with Department of Electrical, Computer, and Biomedical Engineering, Toronto Metropolitan University, 350 Victoria St., Toronto, Ontario, M5B 2K3, (mahdi.shamsi@ryerson.ca, soosan@ryerson.ca) }
}



\maketitle

\begin{abstract}
This paper proposes an impulse response modeling in presence of input and noisy output of a linear time-invariant (LTI) system. The approach utilizes Relative Entropy (RE) to choose the optimum impulse response estimate, optimum time delay and optimum impulse response length. The desired RE is the Kulback-Lielber divergence of the estimated distribution from its unknown true distribution. A unique probabilistic validation approach estimates the desired relative entropy and minimizes this criterion to provide the impulse response estimate. Classical methods have approached this system modeling problem from two separate angles for the time delay estimation and for the order selection. Time delay methods focus on time delay estimate minimizing various proposed criteria, while the existing order selection approaches choose the optimum impulse response length based on their proposed criteria. The strength of the proposed RE based method is in using the RE based criterion to estimate both the time delay and impulse response length simultaneously. In addition, estimation of the noise variance, when the Signal to Noise Ratio (SNR) is unknown is also concurrent and is based on optimizing the same RE based criterion.

The RE based approach is also extended for online impulse response estimations. The online method reduces the model estimation computational complexity upon the arrival of a new sample. The introduced efficient stopping criteria for this online approaches is extremely valuable in practical applications. Simulation results illustrate precision and efficiency of the proposed method compared to the conventional time delay or order selection approaches. Not only RE based method outperforms these approaches, but also is shown to be more robust to the variations of noise signal to noise ratio (SNR). The results also illustrates the role of the data length and the SNR in this type of data based LTI system modeling.
\end{abstract}


\section{Introduction}

 Linear Time-Invariant (LTI) systems characterize a wide range of dynamics around us. Modeling impulse response of  these systems by the use of a finite length input and noisy output is the focus of this work. In practical applications, due to the uncertainty caused by the noisy observation, issues such as  underparamatrization or overparametrization of the impulse response estimate cause very challenging problems. Most of the existing estimators use the mean square error (MSE) to find the parameter estimates in this setting. 
 If the data has length of $N$, it is known that up to $N$ coefficients of the impulse response can be estimated by the available data \cite{Beheshti2}. However, 
 the main challenge in this scenario is which subspace of these coefficients should be chosen for the MSE estimate. In this paper we rely on Relative Entropy (RE) to find the optimum estimate of the impulse response.  Relative Entropy (also known as Kullback–Leibler divergence) is a statistical distance that measures the difference between two probability distributions. Relative entropy is a divergence in terms of information geometry \cite{amari2016information} and has wide a range of applications from theory of information theory to signal processing \cite{lai2009relative}, sensor networks \cite{tang2018information}, cryptography \cite{bose2008information}, machine learning \cite{mackay2003information, seghouane2004small} and physics \cite{lesne2014shannon}. If the relative entropy between two probability distributions is zero, then these two distributions have identical quantities of information \cite{techakesari2013relative}. In this system modeling context, relative entropy can quantify the similarity between the unavailable model and the approximated model.
The proposed RE based method provides the  impulse response coefficients estimate and includes the optimum estimate of the associated time  delay as well as the optimum estimate of the impulse response length. 

Order selection or estimation of the impulse response length is an important task for the purpose of overestimation avoidance \cite{Beheshti2} and model selection methods are critical in a wide range of applications and in areas such as brain source localization \cite{jatoi2014eeg}, wireless sensor networks \cite{le2007adaptive} and machine learning  \cite{bartlett2002model}. While the existing order selection methods concentrate on estimating the length of the impulse response, they don't estimate the time delay of the system. On the other hand time delay estimation methods such as Cumulative Sum (CUSUM) method \cite{gustafsson2000adaptive}, frequency domain based method \cite{Falk} and parametric methods (MATLAB\textregistered \textit{delayest}) use different criteria to estimate the time delay only. Note that time delay itself is an ubiquitous physical phenomenon which often occurs in communication systems \cite{carrasco2012time, he2019partial}, power systems \cite{Milan1}, biological systems \cite{Iones1,laguana}, transportation systems, mechatronic systems \cite{kim2019} and industrial processes such as chemical processing systems \cite{Mehr1}. Inaccurate or improper time delay estimation can cause deterioration in stability and performance in control process. Estimating the time delay is necessary in system modeling, identification algorithms and control systems \cite{Alves1} \cite{Richard1}. 
 
The proposed RE based method provides the optimum impulse response with the optimum length and delay all at once by minimizing the desired relative entropy criterion. Through a unique  approach  probabilistic bounds on the Relative Entropy are calculated which enables this optimization procedure. The powerful relative entropy criterion even enables the method to simultaneously estimate the variance of the output noise when it is unknown. In this procedure, estimation of the reconstruction error, which is the mean squared error (MSE) between the estimated output and the unavailable noise-free output, is required.
Preliminary work for estimation of this error is available \cite{Beheshti1, Beheshti2}. The probabilistic bounds on reconstruction error have shown to be useful and efficient in a wide range of applications such as blind source separation \cite{naghsh2022unified}, brain source localization \cite{sadat2021efficient}, compressed sensing \cite{rezaii2018ecg} and number of source signal estimation \cite{beheshti2018number}. These bounds are shown to be also important and relevant in calculation of the desired RE criterion.  

In many real world applications of impulse response estimation such as sensor validation \cite{brooks2018sensor} and power systems \cite{carlini2021online}, it is desirable to estimate the impulse response of the system, online. For example, Power systems need to respond quickly to the power outages to avoid generators damage, under-frequency load-shedding, and fault cascades \cite{carlini2021online}. Therefore, online monitoring of power systems is a crucial element in stability of power networks. In general, in online applications the dimension of the data increases rapidly as time grows. Consequently the computational complexity and therefore it's cost, increases accordingly \cite{lin1999time, moon2008universal} and re-estimating the optimal impulse response with arrival of the new data is not efficient. To address this problem we equip the proposed RE based method with a recursive online estimation strategy. The method updates the time delay and model order estimates with respect to the new received data. As a result the complexity order of the impulse response estimation is decreases from $O(N^3)$ to $O(N^2)$. Furthermore, a novel efficient stopping criterion for this online system modeling is introduced in this system modeling. The stopping criterion refers to the condition that must be met in order to terminate the execution of the algorithms. Therefore, it is an important factor in efficiency of the online estimation methods \cite{asefi2020distributed}.  It is important to note that this online estimation, empowered with the consistent stopping criterion, has a great potential as a reliable method for modeling slowly varying LTI systems. 

The paper is organized as follows: Section \ref{section2} states the problem and Section \ref{section3} introduces the mathematical foundation and notations used in this paper. Relative Entropy RE-based system modeling is presented in Section \ref{section4}. Section \ref{section5} introduces the online RE-based system modeling. Simulation results are provided in Section \ref{section6} and Section \ref{sec:conclusion} is the conclusion.

\section{Problem Statement}
\label{section2}
Consider a causal linear time-invariant (LTI), single-input single-output system with the following noise free output
\begin{eqnarray}
\begin{aligned}
\bar y(n)=\sum^{\infty}_{i=0} \bar \theta(n)u(n-i)  \label{equ1}
\end{aligned}
\end{eqnarray}
where $u(n)$ is the input at time $n$, and $\bar \theta(n)$ is the unknown impulse response of the system which can be represented in the following column vector format
\begin{eqnarray}
\bar \theta&=& [\bar \theta(0),\;\cdots, \; \bar \theta(\bar{d}), \; \dots ] \nonumber\\
&=&[\underbrace{0, \; \dots, \; 0}_{\mathclap{\bar d}}, \; \bar \theta(\bar{d}), \; \bar \theta(\bar{d}+1), \; \dots ] \label{tIRC}
\end{eqnarray}
where $\bar d$ is the true unknown delay of the system.
 The observed noisy output of the system is  
\begin{eqnarray}
y(n)=\bar y(n) + \omega(n)   \label{sys1}
\end{eqnarray}
where $\omega(n)$ is the additive white Gaussian noise with zero mean and variance of $\sigma_\omega^2$.

Given the following input-output data with length $N$,
\begin{eqnarray}
u^N &= &[u(0), \; u(1), \dots,  u(N-1)] \nonumber\\ y^N &=& [y(0), y(1), \dots, y(N-1)] \label{uyN}
\end{eqnarray}
the goal is to estimate the impulse response and the corresponding time delay, $\bar d$, in (\ref{tIRC}). In this estimation the optimum length of the impulse response has to be simultaneously provided. Note that this problem setting is different from problems such as echo cancellation in which no impulse response is involved and signal delay is estimated based on comparing two noisy observations \cite{so1}.
\section{Impulse response Mean Square Error (MSE) Estimate}
\label{section3}
Before proposing the method, the notations are established. Here the true unavailable time delay of the impulse response is $\bar d$ and if the filter is finite length, the unavailable true length is $\bar m$.  First  the MSE estimate of the impulse response for a range of impulse response length $m$ where $1\leq m \leq M$ and for delay $d$,  $1\leq d\leq m$ is calculated. The following will establish notations for the MSE estimate of the impulse response in this scenario.

\subsection{Impulse Response Estimate for time delay $d$}
It is known that given the finite-length, $N$, input-output data, estimate of at most the first $N$ coefficients of the impulse response can be calculated. In practical application the choice of the impulse response length is generalized to $M$ which can be smaller than $N$ to avoid excess noise fitting or due to partial information about the system structure. Consequently, for any chosen value of $M$, such that $M\leq N$, it is desired to find the estimate of the first $M$ coefficients of the unknown impulse response in  (\ref{tIRC}):
\begin{eqnarray}
\bar \theta_{\bar d, M}&=&[\underbrace{0, \; \dots, \; 0}_{\mathclap{\bar d}},\;\underbrace{ \bar \theta(\bar d),\; \bar \theta( \bar  d+1), \; \dots \; \bar \theta (M-1)}_{\mathclap{\bar \theta_{{\bar d,M}}}}  ] \label{bar11}
\end{eqnarray}
where
\begin{eqnarray}
\bar \theta_{{\bar d,M}}&=&[ \bar \theta(\bar d),\; \bar \theta(\bar d+1), \; \dots \; \bar \theta(M-1)  ] 
\end{eqnarray}
denotes the values of $d+1$st to $M$th elements of the impulse response. 

Lets denote all possible first $M$ impulse response coefficients (IRC) with a time delay of $d$ ($0\leq d \leq M$) as follows 

\begin{eqnarray}\label{cir}
 \theta_{{d, M}}&=&[\underbrace{0, \; \dots, \; 0}_{\mathclap{d}},\;\underbrace{ \theta(d),\;  \theta(d+1), \; \dots \;  \theta(M-1)}_{\mathclap{ \theta_{{d,M}}}}  ] 
\end{eqnarray}
\subsection{Impulse Response Estimate for  time delay $d$ and length $m$ }
Even though maximum  value of $M$ length is considered for the impulse response, in real applications the additive noise may  corrupt the MSE impulse response estimate of this length. This noise over fitting can happen  when the true impulse response is of finite length less than $M$, or have infinite length but small values for the tail coefficients such that they are comparable with the noise standard deviation and their estimates are very much corrupted.  In this case we consider class of impulse responses with the following structure ($d<m\leq M$), by generalizing the structure of impulse response coefficients in (\ref{cir}) with the new variable $m$:
\begin{eqnarray}
\label{cir2}
\theta_{d,m}=[ \underbrace{0, \; \dots, \; 0}_{\mathclap{ d \; {\rm zeros}}}, \underbrace{\theta(d), \; \dots, \; \theta(m-1)}_{\theta_{{d,m}}}, \underbrace{0, \; \dots, \; 0}_{\mathclap{M-m \; {\rm zeros}}} ] 
\end{eqnarray}
This view can generalize representation of the true impulse response parameter in (\ref{bar11}) to the following 

\begin{eqnarray}
\label{cir4}
\bar \theta_{\bar d, \bar m}=[ \underbrace{0, \; \dots, \; 0}_{\mathclap{ d\; {\rm zeros}}}, \underbrace{\bar \theta(\bar d), \; \dots, \; \theta(\bar m-1)}_{\bar \theta_{{\bar d,\bar m}}}, \underbrace{0, \; \dots, \; 0}_{\mathclap{M- \bar m \; {\rm zeros}}} ] 
\end{eqnarray}
Note that the $unknown$ true $\bar m$ can be finite and less than $M$ or can be the same as $M$,  specially  in the cases that the true length of the system is either larger than $M$ and even infinity. 

Here the Topelitz matrix generated by the input is
\begin{eqnarray}
A_{{0,N}}&=&
\left [
\begin{array}{cccc}
u(0) & 0 & \dots & 0\\
u({1}) & u(0) & \dots & 0\\
\vdots & \vdots & \ddots & \vdots\\
u(N-1) & u(N-2) & \dots & u(0)
\end{array}
\right ]  \\ & =& [A_{{0,d}} \;\; A_{{d,m}} \;\; A_{{m,M}} \;\; A_{{M,N}}] \label{OSM2}
\end{eqnarray}
Where $A_{{a,b}}$ is a matrix with columns $a+1$st to the $ b$th column of the Toeplitz matrix. Based on (\ref{cir4}) and (\ref{OSM2}), the noise free output of system (\ref{equ1}) is
\begin{eqnarray}
\bar y^N &=& [A_{{0,\bar  d}} \;\; A_{{\bar d,\bar m}}\;\;A_{{\bar m,M}}] \left [
\begin{array}{cccc}
0_{\bar d\times1}\\
\bar \theta_{{{\bar d},\bar m}}\\
0_{(M-\bar m) \times 1}
\end{array}
\right ]  \\
&=&  A_{{0,\bar d}}0_{\bar d\times1}+ A_{{\bar d, \bar m}}\bar \theta_{{{\bar d},\bar m}}+A_{{\bar m,M}}0_{(M- \bar m) \times 1} \nonumber\\
&=&  0\;\;+\underbrace{ A_{{\bar d, \bar m}}\bar \theta_{{{\bar d}, \bar m}}}_{\mathclap{{ \rm delay}\;  \bar d \; {\rm and} \; {\rm length} \; \bar  m }}+\;\;0   \label{17}
\end{eqnarray}
while the observed noisy data (\ref{sys1}) is
\begin{eqnarray}
 y^N &=& [A_{{0, d}} \;\; A_{{ d,m}}\;\;A_{{m,M}}] \left [
\begin{array}{cccc}
\Delta_{{0,d}}\\
 \theta_{{{ d},m}}\\
\Delta_{{m,M}}
\end{array}
\right ] +\omega^N \\
&=&  A_{{0,d}}\Delta_{{0,d}}+\underbrace{A_{{d,m}}  \theta_{{{ d},m}}}_{\mathclap{{ \rm delay}\;  d\; {\rm and}\; {\rm length}  \; m}} +A_{{m,M}}\Delta_{{m,M}}+\omega^N\nonumber\\
\label{18}
\end{eqnarray}
Note that to distinguish the possible unmodeled coefficients of the true parameter, they are denoted by $\Delta$ function as follows
\begin{eqnarray}
\Delta_{{a,b}}=\bar \theta_{{a,b}} \label{delta}
\end{eqnarray}

Impulse response parameters with nonzero values starting at delay $d$ and with the length of $m$ can generate output in the following structure
\begin{eqnarray}
 y^N_{{d,m}} &=& [A_{{0, d}} \;\; A_{{ d,m}}\;\;A_{{m,M}}] \left [
\begin{array}{cccc}
0\\
 \theta_{{{ d},m}}\\
0
\end{array}
\right ]  + \omega^N\\
&=&  A_{{0,d}}\times 0+\underbrace{A_{{d,m}}  \theta_{{{ d},m}}}_{\mathclap{{ \rm delay}\;  d\; {\rm and}\; \;{\rm length} \:  m }} +A_{{m,M}}\times 0 + \omega^N \nonumber\\
\label{18}
\end{eqnarray}
In this case, the MSE estimate of $\theta$  is as follows
 \begin{eqnarray}
 \hat \theta_{{d,m}} &=& \arg \min_{\theta _{{d,m}}}  || y^N - A_{{d,m}} \theta ||^2_2 \nonumber \\
 &=&(A^T_{{d,m}}A_{{d,m}})^{-1}A^T_{{d,m}} y^N
 \label{ircestimate}
 \end{eqnarray}
Using the structure in (\ref{cir2}) this estimate can equivalently be represented as
\begin{eqnarray}
\label{cir3}
\hat \theta_{d,m}=[ \underbrace{0, \; \dots, \; 0}_{\mathclap{ d}},\hat \theta_{{d,m}}, \underbrace{0, \; \dots, \; 0}_{\mathclap{M-m}} ] 
\end{eqnarray}
 
The estimate of the observed output, according to the estimated IRC is
\begin{eqnarray}
\hat y^N_{{d,m}}&=& A_{{d,m}} \hat \theta_{{d,m}} 
\label{yhat}
\end{eqnarray}
which using  (\ref{cir3}) and (\ref{18})  is equivalently
\begin{eqnarray}
\hat y^N_{{d,m}}&=& A_{{0,M}} \hat \theta_{{d,m}} 
\label{yhat_2}
\end{eqnarray}

\subsection{Summary of Notations and the following Question}
$y^N, u^N$ in (\ref{uyN}): Input and output have fixed length $N$.

$\theta$ in (\ref{cir2}), $\bar \theta$ in (\ref{cir4}): While $\theta$ represents impulse response parameters if there is a bar over the $\theta$, $\bar \theta$ is the true unavailable parameter.

$\hat \theta$, $\hat\theta_{d,m}$ in (\ref{ircestimate}): If there is a hat over $\theta$,  $\hat \theta$ is an estimate of the impulse response. In addition $\theta_{{d,m}}$  are values of the $\theta$ from $d+1$ to $m$ itself with length of $m-d$. The MSE estimate of $\theta$ for each delay $d$ and with total length of $m$ is $\hat \theta_{d,m}$.   
 
$\hat y^N_{d,m}$ in (\ref{yhat}), $\bar y^N$ in (\ref{17}):  The parameter estimate $\hat \theta_{d,m}$ consequently produces an estimate of the output that is denoted by $\hat y^N_{d,m}$. While $\bar y^N$ is the noise free output of the system, $\hat y^N_{d,m}$ is the estimate of this value using the available noisy data.

Next the goal is to compare $\hat \theta _{d,m}$s,  estimates of the system, for a range of delay $d$ and impulse response length $m$, and choose the one that optimally represents the unknown system. In the following section the proposed relative entropy criterion is calculated for the purpose of this comparison.   

\section{Relative Entropy (RE)-Based LTI  modeling}
\label{section4}
The proposed method in this section determines the optimum impulse response estimate based on relative entropy criterion. The true unknown impulse response has the unknown delay of $\bar d$ and unknown length of $\bar m$ (can be infinity).

From (\ref{equ1}) and (\ref{sys1}) the probability distribution function of the observed output given the true parameter $\bar \theta$ and input $u^N$ is

\begin{align}
    f_y(y^N;\bar \theta,u^N) = \frac{1}{(\sqrt{2 \pi \sigma^2_\omega})^N} e^{{-\frac{||y^N - \bar y^N||_2^2}{2 \sigma^2_\omega}}} \label{eq:fx}
\end{align}
which is a Gaussian distribution with mean $\bar y$ in (\ref{sys1}). 
On the other hand, probability distribution function of the output given the estimated parameter $\hat \theta_{d,m}$, with delay $d$ and length $m$ (\ref{ircestimate}), and  input $u^N$ is:

\begin{align}
    g_y(y^N;\hat \theta_{d,m},u^N) = \frac{1}{(\sqrt{2 \pi \sigma^2_\omega})^N} e^{-\frac{||y^N - \hat y^N_{d,m}||_2^2}{2 \sigma_\omega^2}} \label{eq:gx}
\end{align}

where $\hat y^N_{{d,m}}$ is generated by the estimated $\hat \theta_{d,m}$ in (\ref{yhat_2}). This is a Gaussian distribution with mean $\hat y^N_{{d,m}}$. 

It is known that the relative entropy between two multi variant normal distributions $f(y^N)$, with mean $\mu_1$ and covariance matrix $\Sigma_1$, and $g(y^N)$, with mean $\mu_2$ and covariance matrix $\Sigma_2$, is \cite{hershey2007approximating, dytso2019mmse}

\begin{align}
    D(f||g) &= \frac{1}{2}(\log\frac{|\Sigma_2|}{|\Sigma_1|} - N+ \nonumber \\
&\text{tr} \{ \Sigma_2^{-1}\Sigma_1 \} + (\mu_2 - \mu_1)^T \Sigma_2^{-1}(\mu_2 - \mu_1)) \label{eq:NKL}
\end{align}

The relative entropy between the true distribution of the output in (\ref{eq:fx}) and the estimated distribution in (\ref{eq:gx}) (equivalently denoted by $g_{d,m}$) based on (\ref{eq:NKL}) is

\begin{align}
    D(f || g_{d,m}) &= \frac{1}{2}( \log \frac{ \sigma^2_\omega}{\sigma^2_\omega} -N + \frac{\sigma^2_\omega}{ \sigma^2_\omega} + \frac{||\bar y^N - \hat y^N_{d,m} |2^2}{ \sigma^2_\omega}) \\
    &=\frac{1}{2}( 1 -N + N\frac{z_{d,m}}{ \sigma^2_\omega})\label{DLfg}
\end{align}
where $z_{{d,m}}$ is
the distance between the true $unavailable$ noise free output $\bar y^N$ and the estimated output $\hat y_{d,m}$ and is denoted as the \textbf{Reconstruction Error}:
\begin{eqnarray}
z_{{d,m}}= \frac{1}{N}|| \bar y^N - \hat y_{{d,m}}^N||^2_2  \label{zsm}
\end{eqnarray}
Now the goal is to compare the relative entropy of the impulse response estimates of different delay and length and to chose the optimum estimate $\hat \theta_{d^*,m^*}$ with the optimum delay $d^*$ and optimum length $m^*$ such that this criterion is minimized,

As (\ref{DLfg}) shows, minimizing the relative entropy is
equivalent to minimizing the reconstruction error $z_{d,m}$.   
While the true noise free output is not available, probabilistic bounds on the reconstruction error are provided by using the connection of the mean and variance of this random variable with the available exponential term in (\ref{eq:gx}), $|| y^N - \hat y^N_{d,m} ||_2^2$  . Lets denote the \textbf{output error} with the following notation:
\begin{eqnarray}
x_{{d,m}} &=& \frac{1}{N}|| y^N - \hat y^N_{{d,m}}||^2_2  \label{xsdmnew}
\end{eqnarray}
For each $m$ and $d$ a sample of this random variable is available. This one sample will help to provide probabilistic bounds on the mean and variance of $z_{{d,m}}$ as shown in the following section.

\subsection{Probabilistic Estimation of Reconstruction Error}
Mean and variance of the output error and mean and variance of the reconstruction error are provided in the following Lemma: 

{\em Lemma 1:}
The output error, $x_{{d,m}}$, defined in (\ref{xsdmnew}) , is a sample of Chi-square random variable, $X_{{d,m}}$, and the reconstruction error $z_{{d,m}}$, defined in (\ref{zsm}), is a sample of Chi-squared random variable $Z_{{d,m}}$ with the following expectations and variances 
\begin{align}
E(X_{{d,m}}) &=(1-\frac{m-d}{N})\sigma_w^2+ {\Delta_{d,m}} \;\;\;\;\;\;\;\;\;\;\;\;\;\;\;\; \\
{\rm var}(X_{{d,m}}) &=\frac{2}{N}(1-\frac{m-d}{N})(\sigma_w^2)^2+ \frac{4\sigma_w^2}{N}{\Delta_{d,m}}\label{eqn:mine22}\\
{\rm E}(Z_{{d,m}}) &= \frac{m-d}{N}\sigma_w^2+{\Delta_{d,m}}  \label{eqn:aroos2} 
\\ {\rm var}(Z_{{d,m}}) &= \frac{2(m-d)}{N^2}(\sigma_w^2)^2 \;\;
\label{eqn:doomad222}
\end{align}

where $\sigma_w^2$ is the additive noise variance in (\ref{sys1}) and  $\Delta_{d,m}=\frac{1}{N}||G_{{d,m}}F_{d,m}||_2^2$, with $G_{d,m}$ and $F_{d,m}$ defined as follows:

\begin{eqnarray}
G_{{d,m}}=I-A_{{d,m}}(A_{{d,m}}^TA_{{d,m}})^{-1}A_{{d,m}}^T \\
{F_{d,m}} = \left[ {\begin{array}{*{20}{c}}
{{A_{{{0,d}}}}}&{{A_{{{m,M}}}}}
\end{array}} \right]\left[ {\begin{array}{*{20}{c}}
{{\Delta _{{{0,d}}}}}\\
{{\Delta _{{{m,M}}}}}
\end{array}} \right] \label{fnew}
\end{eqnarray}

{\em Proof}: In Appendix A.\\

Consequently, for each $m$ and $d$, calculation of the expected value and mean of the reconstruction error requires knowledge of noise variance and the additional term $\Delta_{d,m}=\frac{1}{N}||G_{{d,m}}F_{d,m}||_2^2$. Note that $\Delta_{d,m}=\frac{1}{N}||G_{{d,m}}F_{d,m}||_2^2$ can be zero or nonzero due to the possible unmodeled dynamics in the subspace with delay $d$ and length $m$.

Using the available sample of the output error, the following theorem provides probabilistic worst case bounds on the reconstruction error:\\ 

{\em Theorem 1:}
Using the available calculated output error $x_{{d,m}}$,  in (\ref{xsdmnew}) , the upper bound and the lower bound of the reconstruction error, in (\ref{zsm})),  with confidence probability $Q(\beta)$  and validation probability $Q(\alpha)$ are 
\begin{eqnarray}
\overline{z_{{d,m}}}= U_{{d,m}}+ \frac{m-d}{N}\sigma_w^2+\beta\frac{\sqrt{2(m-d)}\sigma_w^2}{N} \label{eqn:aziz} \\
\underline{z_{{{d,m}}}}=\max\{0, L_{{d,m}}+\frac{m-d}{N}\sigma_w^2- \nonumber \;\;\;\;\;\;\;\;\;\;\;\;\; \\ \beta\frac{\sqrt{2(m-d)}\sigma_w^2}{N}\} \label{zsmunew}
\end{eqnarray}

where
\begin{eqnarray}
U_{{d,m}}=x_{{d,m}}-c_{d,m}+\frac{2\alpha^2\sigma_w^2}{N}+ \kappa_{{d,m}}(\alpha)\label{eqn:unew} \\
L_{{d,m}}=x_{{d,m}}-c_{d,m}+\frac{2\alpha^2\sigma_w^2}{N}- \kappa_{{d,m}}(\alpha)
\label{eqn:lnew}
\end{eqnarray} 
and
\begin{eqnarray}
\kappa_{{d,m}}(\alpha)&=&2\alpha
\frac{\sigma_w}{\sqrt{N}}\sqrt{\frac{\alpha^2\sigma_w^2}{N}+
x_{{d,m}}-\frac{1}{2}c_{d,m}}\\
c_{d,m}&=&(1-\frac{m-d}{N})\sigma_w^2.
\end{eqnarray}

where $Q(\alpha) = \int_{-\alpha}^{\alpha}\frac{1}{\sqrt{2\pi}}e^{-x^2/2}dx$.\\ 

Proof: In Appendix B.\\

The upper bound of the reconstruction error is the worst case probabilistic upper bound that can provide the optimum value of the delay and length with respect to the relative entropy criterion (\ref{DLfg})
\begin{align}
 (d^*, m^*) &= \mathop {\arg \min }\limits_{d,m} \overline{D(f || g_{d,m})} = \nonumber \\ & \mathop {\arg \min }\limits_{d,m} \frac{1}{2}( 1 -N + N\frac{\overline{z_{d,m}}}{ \sigma^2_\omega})
\end{align}

where $\overline{D(f || g_{d,m})}$ is the calculated probabilistic worst case of the desired relative entropy.  

In the above calculation the noise variance is assumed to be known. In the following section we optimize the relative entropy for the case of unknown noise variance. 

 \subsection{Relative Entropy Criterion with Unknown Noise Variance}\label{noisevar7}

 Unknown noise variance $\sigma^2_w$ in the desired RE criterion in (\ref{DLfg}) can be treated similar to the unknown time delay $\bar d$ and impulse response length $\bar m$. Similarly a range of possible noise variance values between $\sigma_{min}$ and $\sigma_{max}$, is considered 
 $\sigma \in [\sigma_{min}, \cdots, \sigma_{max}]$. In this case the true probability distribution of the observed data in (\ref{eq:fx}) can be one of the following distributions:
 
 \begin{align}
    f(\sigma) = \frac{1}{(\sqrt{2 \pi \sigma^2})^N} e^{{-\frac{||y^N - \bar y^N||_2^2}{2 \sigma^2}}} 
\end{align}
 while the distribution of the data based on the estimated output is
 
 \begin{align}
    g(y^N;\hat \theta_{d,m},u^N,\sigma) = \frac{1}{(\sqrt{2 \pi \sigma^2})^N} e^{-\frac{||y^N - \hat y^N_{d,m}||_2^2}{2 \sigma^2}} \label{eq:gx}
\end{align}
 
The upperbound of the reconstruction error in (\ref{eqn:aziz}) can be calculated for each of these noise variances denoted by
 $\overline{z_{{d,m,\sigma}}}$. Therefore, the relative entropy associated to these $\sigma$s can be calculated as follows (\ref{DLfg}):

\begin{align}
      \overline{D_{d,m,\sigma}(f(\sigma)||g_{d,m}(\sigma))} =\frac{1}{2}(1 - N + N \frac{\overline{z_{{d,m,\sigma}}}}{\sigma^2}) \label{eqn:dlfinal}
 \end{align}

Minimizing the relative entropy estimate between $f(\sigma)$ and $g_{d,m}(\sigma)$ determines the optimum values $d^*$, $m^*$ and the optimum noise variance ${\sigma^*_\omega}^2$

\begin{eqnarray}
(d^*,m^*, {\sigma^*_\omega}^2 )=\arg \min_{d,m,\sigma} \overline{ D_{d,m,\sigma}(f(\sigma)||g_{d,m}(\sigma))} \label{godd}
\end{eqnarray}

Algorithm \ref{Alg_1} shows the complete pseudo code of the proposed RE based impulse response estimation method.

\begin{algorithm}[ht!]
	\small
	\centering
	\caption{RE based IR estimation with optimum delay $d^*$ and length $m^*$ and noise variance ${\sigma^*_\omega}^2$}
	\label{Alg_1}
	\begin{algorithmic}[1]
		\REQUIRE Input and output data $ {x^{N}} = [x_1,x_2,...,x_N] $ and ${y^N}=[y_1,y_2,...,y_N]$, range for noise variance $\sigma \in [\sigma_1, \sigma_2, \sigma_3, \dots, \sigma_K]$, $\alpha$ and $\beta$, maximum length of the IRCs $M$
		\ENSURE Estimated time delay $d^*$, length of the IRCs $m^*$ and estimated noise variance $\sigma^*_\omega$ 
		\FOR{$i=1; i \leq K;i_{++}$}
		\STATE $\sigma = \sigma_i$
		\FOR{($m=1; m \leq M;m_{++}$)}
		\FOR{($d=1; d \leq m;m_{++}$)}     	 
		\STATE Estimate the IRCs $\hat \theta_{{d,m}}$ based on (\ref{ircestimate})
		\STATE Calculate the output error, $x_{{d,m}}$ from (\ref{xsdmnew})
		\STATE Calculate the upper bound of reconstruction error $\overline{z_{{d,m,\sigma}}}$ from (\ref{zsmunew}) to (\ref{eqn:unew})
		\IF{$\overline{z_{{d,m,\sigma}}} \in \mathbb{R}^+$}
		\STATE continue the algorithm
		\ELSE
		\STATE go to next value of $\sigma$
		\ENDIF
		\ENDFOR
		\ENDFOR
		\ENDFOR
		\IF{All the calculated $\overline{z_{{d,m}}}\in \mathbb{R}^+$}
		\STATE Estimate the optimum time delay $d^*$ and length of IRCs $m^*$ and estimated noise variance ${\sigma^*_\omega}^2$ (\ref{godd}): \\
		$(d^*,m^*, {\sigma^*_\omega}^2 )=\arg \min_{d,m,\sigma} \overline{ D_{d,m,\sigma}(f(\sigma)||g_{d,m}(\sigma))}$
		\ENDIF		
	\end{algorithmic}
\end{algorithm}

 Note that calculation of each of the IR estimate $\hat \theta^N_{d,m}$
is of order $O(N^3)$ due to the use of Toeplitz matrix and the use of inverse matrices. On the other hand the process of RE minimization will not add any complexities of the order of the data length as the range of possible delay and possible impulse response length are chosen as finite values. 
\section{Online RE based Impulse Response (IR) Estimation with Optimum Delay and Length}
\label{section5}
In some practical applications the received observations are sequential and each time a new data point arrives and the data set is updated. This additional information should then be used in recomputing the parameter estimates of the system and update the existing parameter estimates with respect to this new received data point. Recomputing the impulse response estimate from scratch is costly and requires exponential computing time and memory. Therefore, for online practical applications it is desirable to update the existing parameter estimate recursively. Let's denote the least square estimate of the impulse response of the system after receiving $N$ observations by $\hat{\theta}^N$. The updated impulse response estimation after receiving the $(N+1)^{\mathop{th}}$ observation
(\ref{cir3}) is $\hat \theta_{d,m}^{N + 1}$:
\begin{eqnarray}
\hat \theta_{d,m}^N \to \hat \theta_{d,m}^{N + 1}
\end{eqnarray}
and consequently the output error (\ref{xsdmnew}) as a function of $N$ is updated to:
  \begin{eqnarray}
  x^{N+1}_{{d,m}} =  \frac{1}{N+1}|| y^{N+1} - \hat y^{N+1}_{{d,m}}||^2_2  \label{xsN}
  \end{eqnarray}
which will lead to an update of the upperbound on the reconstruction error that itself leads to updates of the estimated delay and and estimated impulse response length:
\begin{eqnarray}
d^*_N,m^*_N \to d^*_{N + 1},m^*_{N + 1}
\end{eqnarray}
The following subsection provides the details of the online recursive procedure for the proposed RE optimization method. In addition, note that for practical online applications an stopping criterion is essential and valuable. Next we propose an stopping criterion for the procedure as $N$ grows. The stopping criterion is beneficial and crucial in practical applications and will also help expanding application of the RE based method for modeling systems with coefficients that are gradually changing with the time.   
\subsection{Updating Procedure}
The parameter estimate (\ref{ircestimate}) using data length of $N+1$ is 
\begin{eqnarray}
 {\hat\theta_{{d,m}}^{N+1}} =  \mathop {\arg \min }\limits_{\theta \in {d,m}} || y^{N+1} - A_{{d,m}}^{N+1} \theta ||^2_2  \\
 =((A_{{d,m}}^{N+1})^TA^{N+1}_{{d,m}})^{-1}{(A^{N+1}_{{d,m}})}^T y^{N+1}
 \label{ircestimateN}
 \end{eqnarray}
 where $A_{{d,m}}^N$ is the columns of Toeplitz matrix in (\ref{OSM2}) from index $d+1$ to index $m$. If the data length increases by one, a new column and a new row are added to the matrix in (\ref{OSM2}) and consequently we have
With the  $N+1$th data we have
\begin{eqnarray}
\label{eqn:ANPlus}
A_{{d,m}}^{N+1}=\left [
\begin{array}{c}
A_{{d,m}}^N\\
(B_{d,m}^N)^T 
\end{array}
\right ]
\end{eqnarray}
where 
\begin{multline}
(B_{d,m}^N)^T = \\
\left[ {\begin{array}{*{20}{c}}
{{u((N + 1) - (d + 1))}}{\begin{array}{*{20}{c}}
 \cdots {{u((N + 1) - m)}}
\end{array}}
\end{array}} \right]
 \end{multline}
 
The following Lemma uses the available $\hat \theta^N_{d,m}$ and $u(N+1)$ and $y(N+1) $ to provide $\hat\theta_{{d,m}}^{N+1}$

 {\em Lemma 3:}
 Recursive method of updating $\hat \theta_{d,m}^N \to \hat \theta_{d,m}^{N + 1}$ is as follows:
 \begin{eqnarray}
 {\hat\theta_{{d,m}}^{N+1}}=(K^N_{{d,m}})^{ - 1} - 
   \frac{1}{1 + {\rm tr}(\gamma_{d,m}^N)}C_{{d,m}}^N(C_{{d,m}}^N)^T \times \nonumber \\
 (({A_{{d,m}}^N})^Ty^N+B_{d,m}y(N+1)) 
\\
 =(I-\frac{1}{1 + {\rm tr}(\gamma_{d,m}^N)}\gamma_{{d,m}}^N)\hat \theta _{{{d,m}}}^{N}+ \nonumber\\
 (I+\gamma_{d,m}^N)C_{{d,m}}^Ny(N+1)
 \end{eqnarray}
 
 \noindent where 
 
 \begin{align}
&C_{{d,m}}^N=(K^N_{{d,m}})^{ - 1}B_{d,m}^N\\
&\gamma_{d,m}^N=(K^N_{{d,m}})^{ - 1}B_{d,m}^N(B_{d,m}^N)^T
 \end{align}

\noindent and tr$(a)$ is trace of matrix $a$.

As a result of this recursive calculation, the complexity order of non recursive calculation of $\hat \theta_{d,m}^{N+1}$ which is of order $O(N^3)$ will be reduced to $O(N^2)$.\\ 

Proof: In Appendix C.\\
 
This parameter update generates updated data error which can be used in calculation of the reconstruction error upperbound in (\ref{zsmunew}) to provide the updated delay and length of the parameter estimate based on (\ref{DLfg}):
 
 \begin{align}
     (d^*_{N+1},m^*_{N+1})=\arg \min_{d,m} \overline{ D_{d,m}(f||g_{d,m})}
 \end{align}
 and chooses the optimum parameter estimates 
 \begin{eqnarray}
 \hat{(\theta^*)}^{(N+1)}=\hat\theta_{{d^*_{N+1},m^*_{N+1}}}^{N+1}
 \end{eqnarray}

\subsection{Stopping Criterion}

For the stopping criterion, the following desired signal to noise (SNR) ratio is logical:

\begin{multline}
    \frac{\mathrm{Best\;Estimate\; of \; the Output}}{\mathrm{Output\;Estimation\;Error}} = \frac{||\hat{y}_{{d^*,m^*}}^{N}||_2^2}{||\bar y^{N} - \hat y_{d^*, m^*}^N||_2^2} =  \\
    \frac{\frac{1}{N}||\hat{y}_{{d,m}}^{N}||_2^2}{{z^{N}_{{d^*,m^*}}}}
\end{multline}
The desired SNR in dB is in the following form and the stopping criterion can be defined based on a desired lower bound on this SNR based on a chosen $\epsilon$
\begin{align}
\label{eqn:epsilonSNR}
    \mathrm{SNR} = 10 \log (\frac{\frac{1}{N}||\hat{y}_{{d,m}}^{N}||_2^2}{{z^{N}_{{d^*,m^*}}}})\;, \;\;\; \mathrm{SNR} > 10\log(\frac{1}{\epsilon})
\end{align}

To utilize such stopping criterion we can use $\overline{z_{{d^*,m^*}}}$ in  (\ref{zsmunew}) which is our probabilistic worse case estimate of the unavailable $z_{d^*,m^*}$. Consequently, the desired stopping criterion is for the first value of $N$ such that the following is satisfied:

\begin{eqnarray}
\frac{\overline {z^{N}_{{d^*,m^*}}}}{\frac{1}{N}||\hat{y}_{{d,m}}^{N}||_2^2} <\epsilon
\label{eqn:Stopping}
\end{eqnarray}
\section{Simulations} 
\label{section6}
We analyze and compare the performance of the proposed Relative Entropy (RE) based system modeling in two scenarios for Finite Impulse Response (FIR) and Infinite Impulse Response (IIR) modeling. Consider the following two time delayed systems:

\textit{System I:} A lowpass FIR filter with  20 kHz pass band-edge frequency, 96 kHz sampling frequency,  0.01 dB peak-to-peak ripple and 80 dB stop band attenuation with time delay $\bar d=7$ and length of $\bar m=69$ \footnote{https://www.mathworks.com/help/dsp/ref/dsp.lowpassfilter-system-object.html}.

\textit{System II:} Non-minimum phase IIR system with the following impulse response \cite{Alves1},
\begin{eqnarray}
\bar \theta(n) = 0.2545(0.9094)^n - 0.3316(0.8146)^n,\;\; n\geq 0
\end{eqnarray}
time delayed by $\bar d=11$.

Both system's Inputs are independent identically distributed (IID) Bernoulli sequence of $\pm 1$ with data length of $N=1000$. 

\subsection{Time Delay Estimation Analysis}
 For time delay estimation, performance of RE based method is compared with three classical methods CUSUM \cite{Alves1}, Frequency-domain and $MATLAB^{\textregistered}$  \textit{delayest}. Delayest method requires the order (number of poles) of the system and this number is by default set to two. In addition, majority of classical time delay estimators are based on thresholding. CUSUM is one of the most used thresholding approaches for time delay estimation and the time delay estimation is highly sensitive to the user selected parameters. For CUSUM approach we use the threshold parameters suggested in \cite{Alves1}. However, this method is based on choosing a maximum value in the frequency domain of the estimated impulse response. This value is highly effected by the noise level even as large as $20$dB in the case of System I. It is worth mentioning that what is shown here as the Frequency-domain method is our improved version compared to its conventional version \cite{Alves1}. In improved version the threshold is further optimized by a trial and error. For RE based method, parameters $\alpha$ and $\beta$ in confidence and validation probabilities are set to 4 as discussed in \cite{Beheshti2} , and therefore, these probabilities are approximately 0.999. Note that the method is robust to changing these values within a wide sufficient and necessary range that is function of data length \cite{Beheshti2}.
 
\begin{figure}[h!]
\centering
\includegraphics[trim=2cm 0.1cm 2cm 0.1cm, scale=0.32]{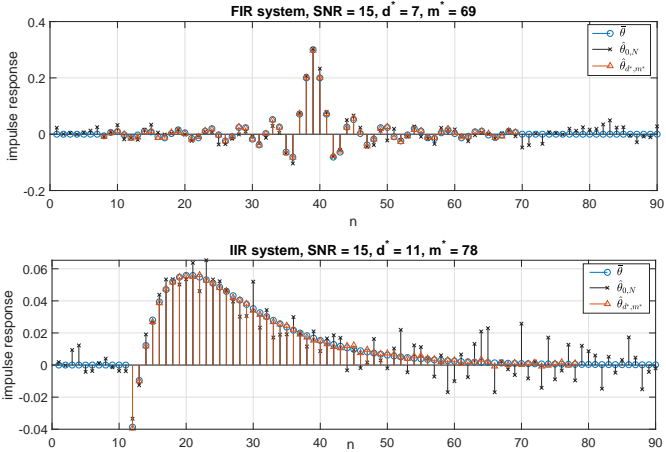}
\caption{ True impulse response coefficients $\bar \theta$, Estimated minimum MSE impulse response of length 1000, $\hat\theta_{0,N}$, Estimated RE based impulse response $\hat\theta_{d^*,m^*}$ }
\label{compare}
\end{figure}

Figure \ref{compare} shows the first 90 coefficients of the two impulse responses in form of $\bar \theta$, (blue signal),  as well as the results of RE based approach for SNR of 15dB. The black signal in the figure also shows two least square estimates of of the coefficients of these impulse responses from zero to 1000, $\hat \theta_{0,N}$, $N=100$. As the data length is 1000, this estimate provides all the 1000 coefficients and as the figure shows these are noisy estimates of the impulse responses. The red signal is $\hat \theta_{d^*,m^*}$,  which is the RE based impulse response estimate. As the figure shows while $\hat \theta_{0,N}$ fits the additive noise, the optimum $d^*$ and $m^*$ for the FIR system are  $d^*=7$ and $m^*=69$ which are the same as the true delay and length ($\bar d=7$, $\bar m=69$). For the IIR system the delay is estimated as $d^*=11$ which is the same as the true unknown delay. Note that although in this case as the impulse response is infinite, ($\bar m= \infty$), the method recognizes that an impulse response with length 78  ($m^*=78$) is the best estimate in this SNR and the rest of coefficients are set to zero to minimize the desired least square of RE.   

\begin{figure}[h!]
\centering
\includegraphics[width=8.5cm]{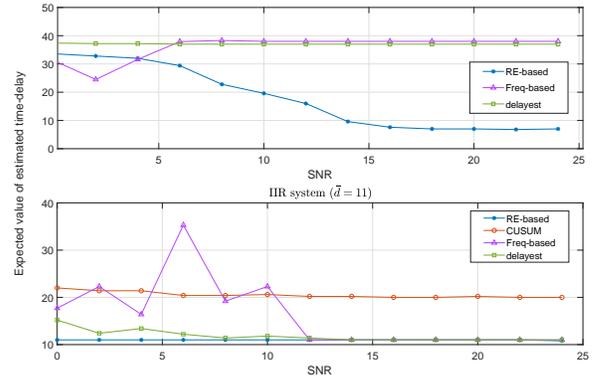}
\caption{Estimated time delay with data length of $N=1000$ for $0 \leq $SNR$ \leq 24$dB (averaged over 100 trials)}
\label{mean1}
\end{figure}

Figure \ref{mean1} shows the average time delay estimate in 100 trials as a function of SNR in dB, where SNR is the ratio of the power of the unavailable noiseless data $\bar y$ and the noise variance $\sigma^2_w$ in (\ref{sys1}). Table \ref{tab1} shows these estimated time delay values as well as their associated standard deviation (SD) in the same 100 trials. As Figure \ref{mean1} and Table \ref{tab1} show, the RE based algorithm outperforms the other three methods. For the FIR case, the time delay estimate is larger than the true time delay for the low SNRs. This is expected as the small values of the first coefficients are comparable with the noise variance for the low SNRs. At around SNR of 15 dB, the first coefficient is comparable with the noise standard deviation and therefore, the method starts choosing the delay as 7 which is the true delay, and the standard deviation of error goes to zero as SNR grows. As the figure shows and the table confirms, the RE based method is the only one that estimates the correct time delay as the SNRs grows. For the IIR system on the other hand, as it is shown in Figure \ref{compare}, the first coefficient of the impulse response has a large absolute value so we expect that an efficient delay estimator chooses the correct time delay even for thelow SNRs. As the table and figure depict, RE based method chooses the correct delay and outperforms the other methods. While delayest is the next method that converges for higher SNRs, Freq-based method also converges for even higher range of SNRs.  

\begin{table*}[ht!]
\begin{center}
\caption{Mean and Standard Deviation (SD) of estimated time delay for $100$ trials of each system}
\label{tab1}
\resizebox{\textwidth}{!}{\begin{tabular}{ | c | c | c | c | c | c | c | c | c || c | c | c | c | c | c | c | c |}
\multicolumn{11}{c}{}\\ \cline{2-17}
\multicolumn{1}{ c|}{} & \multicolumn{8}{| c ||}{} & \multicolumn{8}{| c |}{} \\
\multicolumn{1}{ c|}{} & \multicolumn{8}{| c ||}{\large{FIR System I , ($\bar d=7$)}} & \multicolumn{8}{| c |}{\large{IIR System, ($\bar d=11$)}} \\ \hline

\multicolumn{1}{|c|}{SNR} & \multicolumn{2}{| c |}{RE Based} & \multicolumn{2}{| c |}{CUSUM} & \multicolumn{2}{| c |}{Freq-based} & \multicolumn{2}{| c ||}{Delayest} & \multicolumn{2}{| c |}{RE based} & \multicolumn{2}{| c |}{CUSUM}  & \multicolumn{2}{| c |}{Freq-based} & \multicolumn{2}{| c |}{Delayest}  \\ \cline{2-17}

  (dB)   & mean  &  SD  & mean & SD  & mean & SD & mean & SD  & mean & SD & mean & SD & mean & SD  & mean & SD \\ \hline
   0 & 29.81 & 0.72  & 80.64 &  0.93 & 35.53 & 2.43 & 37.58 & 2.45 & \textbf{11.0} & 0.0 & 15.43 & 0.93  & 22.68 & 12.43 & 14.67 & 2.45  \\ \hline

   2 & 28.72 & 0.69 & 80.29 & 0.71 & 29.10 & 2.57 & 37.53 & 1.61 & \textbf{11.0} & 0.0 & 15.06 & 0.71  & 24.67 & 12.57 & 13.15 & 1.61   \\ \hline

   4 & 23.14 & 0.71 & 80.16 & 0.58 & 32.47 & 1.82 & 37.43 & 1.06 & \textbf{11.0} & 0.0 & 14.73 & 0.58  & 21.99 & 11.82 & 12.74 & 1.06 \\ \hline

   6 & 21.52 & 0.63 & 80.14 & 0.54 & 36.44 & 1.17 &37.39 & 1.04 & \textbf{11.0} & 0.0 & 14.7 & 0.54  & 18.21 & 11.17 & 12.55 & 1.04 \\ \hline

    8 & 18.78 & 0.57 & 80.01 & 0.49 & 36.99 & 0.89 & 37.21 & 0.94 & \textbf{11.0} & 0.0 & 14.61 & 0.49  & 17.49 & 9.89 & 12.53 & 0.94   \\ \hline

   10 & 11.77 & 0.53 & 80.0 & 0.49 & 37.17 & 0.91 &37.11 & 1.00  & \textbf{11.0} & 0.0 & 14.39 & 0.49  & 16.55 & 8.91 & 11.94 & 1.00   \\ \hline

   12 & \textbf{7.34} & 0.41 & 80.0 & 0.46 & 37.93 & 0.29 & 37.10 & 0.86 & \textbf{11.0} & 0.0 & 14.29 & 0.47  & 12.62 & 5.29 & 11.48 & 0.86  \\ \hline

   14 & \textbf{7.2} & 0.37  & 80.0 & 0.46 &37 & 0.13 &37 & 0.13 & \textbf{11.0} & 0.0 & 14.3 & 0.46  & 11.53 & 2.54 & \textbf{11.0} & 0.0 \\ \hline

   16 & \textbf{7.1} & 0.18 & 80.0 & 0.39 & 37.85 & 0.10 & 37.08 & 0.06 & \textbf{11.0} & 0.0 & 14.18 & 0.39  & 11.24 & 1.48 & \textbf{11.0} & 0.0   \\ \hline

   18 & \textbf{7.0} & 0.0 & 80.0 & 0.33 &37 & 0.05 & 37.04 & 0.04 & \textbf{11.0} & 0.0 & 14.12 & 0.32  & 11.01 & 0.05 & \textbf{11.0} & 0.0   \\ \hline

   20 & \textbf{7.0} & 0.0  & 80.0 & 0.24 & 37.48 & 0.23 & 37 & 0.0 & \textbf{11.0} & 0.0 & 14.06 & 0.24  & \textbf{11.0}& 0.0 & \textbf{11.0} & 0.0 \\ \hline

   22 & \textbf{7.0} & 0.0 & 80.0 & 0.1 & 38 & 0.01 & 37 & 0.0 & \textbf{11.0} & 0.0 & 14.01 & 0.1  & \textbf{11.0} & 0.0 & \textbf{11.0} & 0.0   \\ \hline

   24 & \textbf{7.0} & 0.0 & 80.0 & 0.0 & 38 & 0.0 & 37 & 0.0 & \textbf{11.0} & 0.0 & 14.0 &0.0 & \textbf{11.0} & 0.0 & \textbf{11.0} & 0.0   \\ \hline

  \end{tabular}}
 \end{center}
\end{table*}

\begin{figure}[h!]
\centering
\includegraphics[width=8.5cm,height=6.5cm]{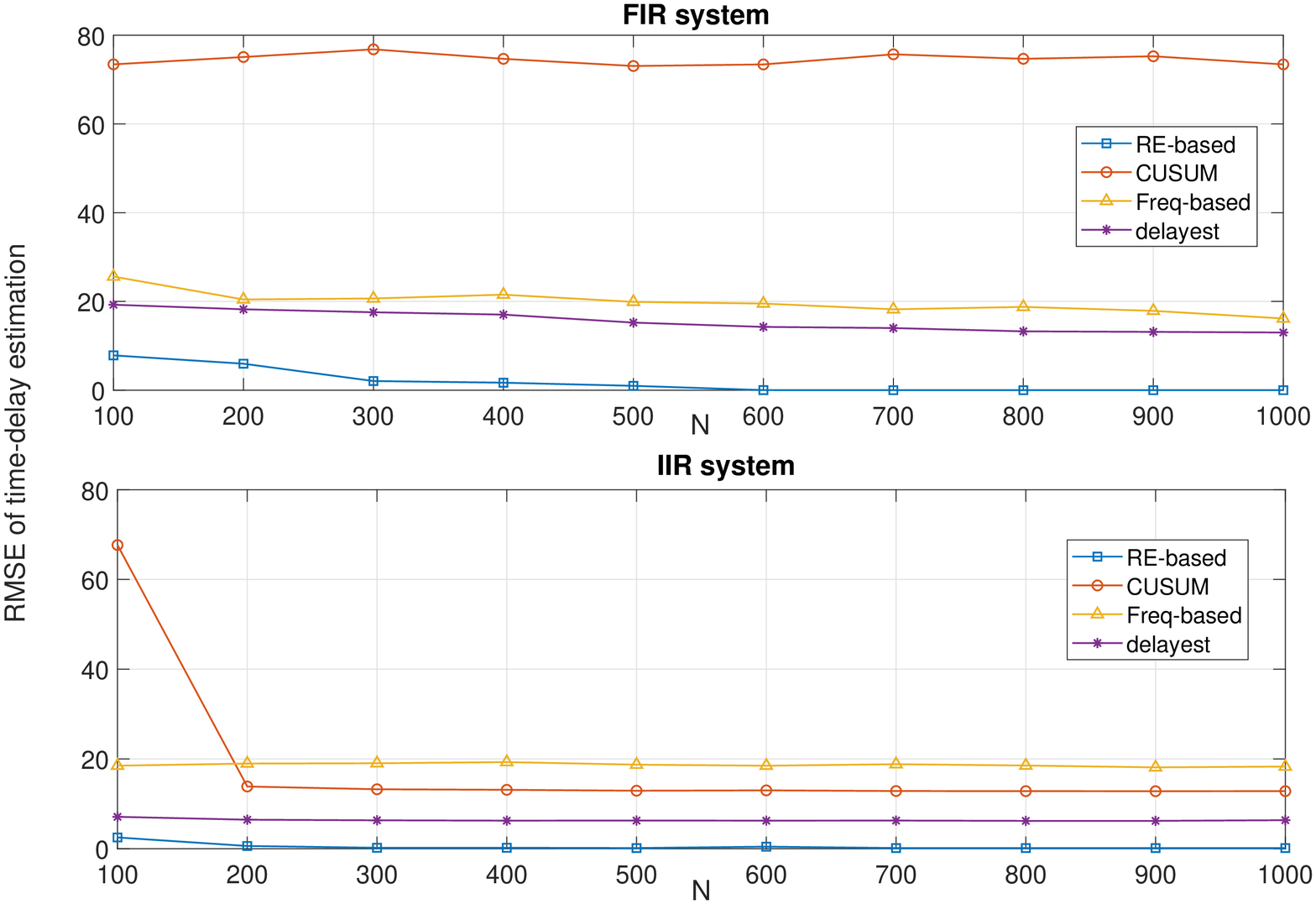}
\caption{Delay RMSE for FIR System and IIR system as the data length $N$ grows, averaged over 100 runs, for randomly generated delays between 1 and 20, for SNR=10db.}
\label{compare4}
\end{figure}

Figure \ref{compare4} shows the RMSE in time delay estimation for the two systems at SNR of 10 dB and as the data length grows from 100 to 1000. In each case, the delay is randomly generated between 1 and 20 (with a uniform distribution) and the RMSE is averaged over 100 runs.  As the figure illustrates RE based method outperforms the other approaches as its error goes to zero as the data length gets around 600 for System I  and around 300 for System II.  While delayest has the next acceptable performance, for none of these method the RMSE approaches zero in this range of data length.

\subsection{Impulse Response Length Estimation Analysis}
\begin{figure}[h!]
\centering
\includegraphics[width=8.5cm]{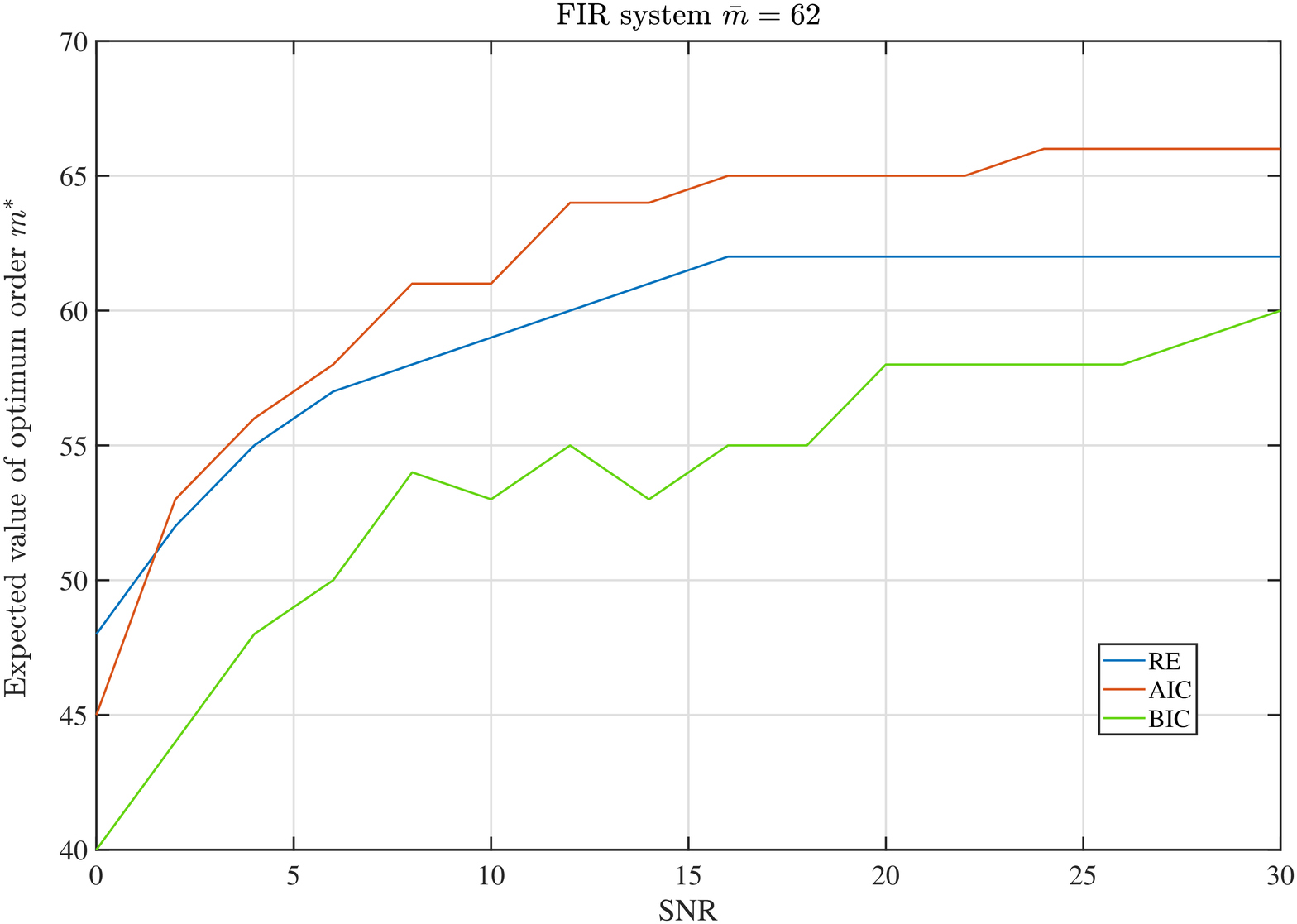}
\caption{Estimated Impulse Length with data length of $N=1000$ for $0 \leq SNR \leq 30$dB (averaged over 100 trials)}
\label{estorder}
\end{figure}

\begin{figure}[h!]
\centering
\includegraphics[width=8cm, trim=2.5cm 0.5cm 2.5cm 0cm]{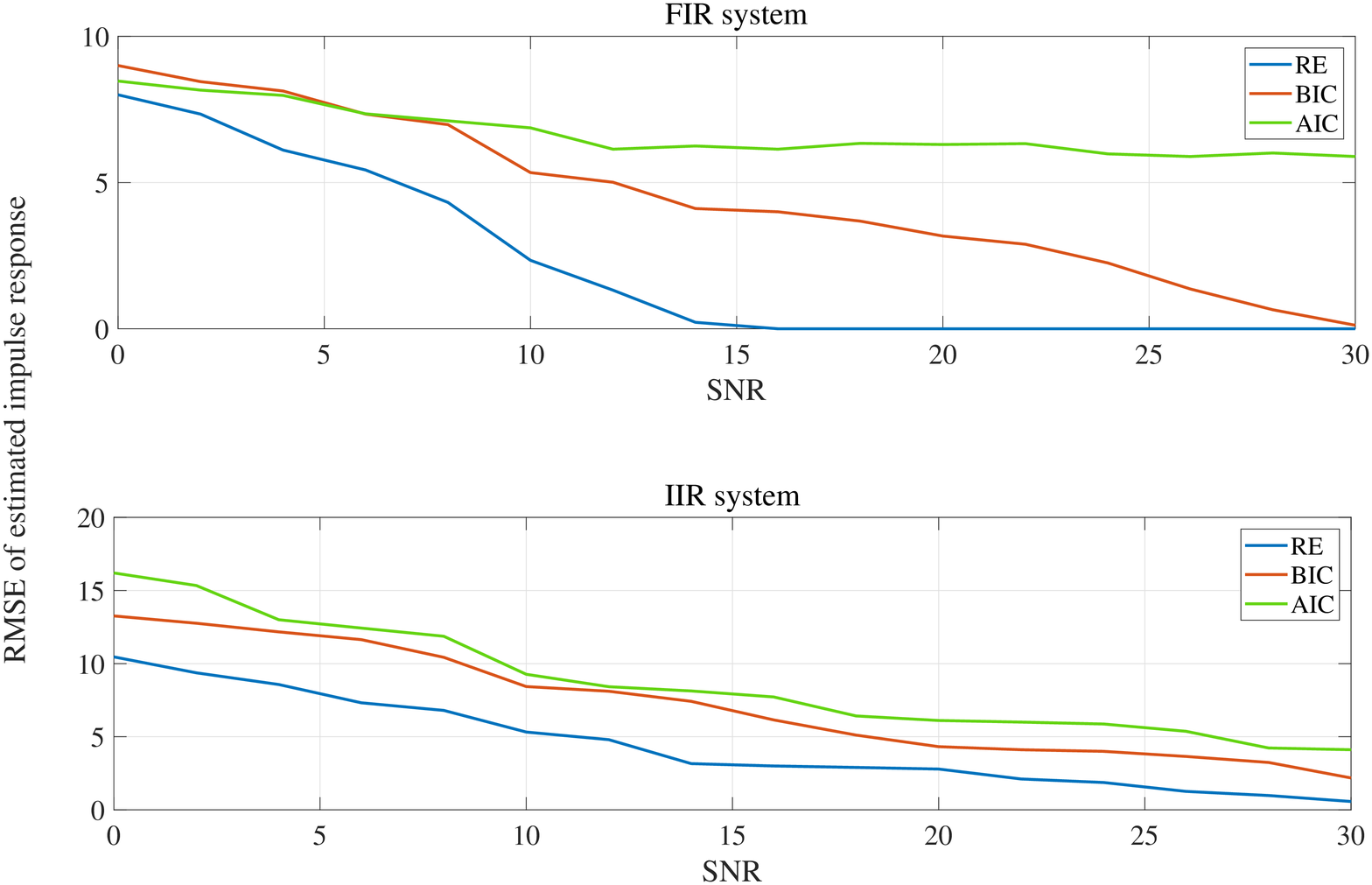}
\caption{RMSE of the estimated impulse response, $N=1000$ for $0 \leq SNR \leq 30$dB for FIR and IIR system averaged over 100 trials}
\label{RMSEm}
\end{figure}

Estimated coefficient length $m$ for a range of SNRs with RE based method is provided in Figure \ref{estorder} for the FIR system ($\bar m =62$). The method is compared with well known and most used order selection methods AIC and BIC. Note that it is shown in \cite{Beheshti2} that AIC and BIC ( equivalently two stage MDL) model order selection methods are special case of the reconstruction error based approach. As the existing order selection methods can't estimate the time delay, we set the time delay to zero for the purpose of comparison. Figure \ref{estorder} shows the simulation results. As the figure shows, for the RE based method, as the SNR grows from zero dB, the chosen impulse response length is growing from 46 points towards the true length of 62 which is chosen after 15 dB. 
As it is expected, the figure shows that the AIC method overestimates, while BIC method under estimates the impulse response length. Figure \ref{RMSEm} shows the RMSE of the estimated impulse response for both FIR and IIR system in the order selection setting and as the SNR grows. As the figure indicates, for the FIR system, RMSE if RE based is minimum of the all method and also for $\mathrm{SNR} > 15 \; \mathrm{dB}$ the RMSE of the proposed RE method approaches zero. While AIC error seems to be nonzero even for SNR of 30dB, the BIC approach converges at this SNR which is much higher than 15dB, that is the convergence point of the proposed RE based method. Note that this simple example shows the important role of SNR in the choice of coefficient lengths and confirms that while the true length of the FIR is 62, it is more efficient to choose less coefficients for lower SNRs and not to fit the additive noise. In other words, what is known as optimum length {\em estimation} for the impulse response is less important than optimum length {\em selection}. As the figure shows RMSE of order selection for the IIR system is also minimum for the RE based method.    

 \subsection{Simultaneous Time Delay Estimation and Impulse Response Length Estimation Analysis}
 In previous sections we compared the RE based method with existing well known time delay estimators and order selection methods. To our knowledge, the proposed RE based method is the only approach that is capable of simultaneous estimation of both the time delay and the impulse response length. Table \ref{tab:final} shows the estimated time delay and impulse response length, as well as the respective RMSE between the true coefficients, $\bar \theta$, and the estimated ones, $\hat \theta_{d^*, m^*}$, when the additive noise variance is unknown. The table shows the results averaged over 100 trials for a Range of SNR. As the table indicates for the FIR system, as the SNR grows, the estimated time delay and impulse response length converge to the true values. For the IIR system, the optimum time delay is estimated correctly for the range of SNR while the estimated optimal impulse response length increases as the SNR grows. This is a rational expectation as the higher the SNR the more valuable are the tails of the least square estimate of the impulse response and therefore the method recognizes to chose more of those estimated coefficients. As the table shows the RMSE in both scenarios become smaller as the SNR grows and is almost zero for SNRs higher than 20dB for the FIR system. 
 
\begin{table}[hbt!]
\centering
\caption{Average estimated time delay, average estimated impulse response length, and the corresponding RMSE for the FIR and IIR systems for a range of SNRs (averaged over 100 trials). }
\label{tab:final}
\begin{tabular}{|c|ccc|ccc|}
\hline
   & \multicolumn{3}{c|}{FIR system ($\bar d= 7$, $\bar m = 69$)}   & \multicolumn{3}{c|}{IIR system ($\bar d = 11$)}              \\ \hline
\begin{tabular}[c]{@{}c@{}}SNR \\ (dB)\end{tabular} &
  \multicolumn{1}{c|}{$d^*$} &
  \multicolumn{1}{c|}{$m^*$} &
  RMSE &
  \multicolumn{1}{c|}{$d^*$} &
  \multicolumn{1}{c|}{$m^*$} &
  RMSE \\ \hline
0  & \multicolumn{1}{c|}{29.81} & \multicolumn{1}{c|}{54.23} & 8.13 & \multicolumn{1}{c|}{11} & \multicolumn{1}{c|}{43.26} & 13.25 \\ \hline
2  & \multicolumn{1}{c|}{28.72} & \multicolumn{1}{c|}{57.16} & 7.26 & \multicolumn{1}{c|}{11} & \multicolumn{1}{c|}{45.18} & 12.64 \\ \hline
4  & \multicolumn{1}{c|}{23.14} & \multicolumn{1}{c|}{62.72} & 6.07 & \multicolumn{1}{c|}{11} & \multicolumn{1}{c|}{50.31} & 11.89 \\ \hline
6  & \multicolumn{1}{c|}{21.52} & \multicolumn{1}{c|}{63.88} & 5.84 & \multicolumn{1}{c|}{11} & \multicolumn{1}{c|}{52.68} & 11.60 \\ \hline
8  & \multicolumn{1}{c|}{18.78} & \multicolumn{1}{c|}{64.47} & 5.21 & \multicolumn{1}{c|}{11} & \multicolumn{1}{c|}{54.45} & 10.85 \\ \hline
10 & \multicolumn{1}{c|}{11.77} & \multicolumn{1}{c|}{66.19} & 4.23 & \multicolumn{1}{c|}{11} & \multicolumn{1}{c|}{59.73} & 10.06 \\ \hline
12 & \multicolumn{1}{c|}{7.34}  & \multicolumn{1}{c|}{66.93} & 3.63 & \multicolumn{1}{c|}{11} & \multicolumn{1}{c|}{63.54} & 8.52  \\ \hline
14 & \multicolumn{1}{c|}{7.2}   & \multicolumn{1}{c|}{68.58} & 1.23 & \multicolumn{1}{c|}{11} & \multicolumn{1}{c|}{65.73} & 7.48  \\ \hline
16 & \multicolumn{1}{c|}{7.1}   & \multicolumn{1}{c|}{69.21} & 0.41 & \multicolumn{1}{c|}{11} & \multicolumn{1}{c|}{70.20} & 6.16  \\ \hline
18 & \multicolumn{1}{c|}{7}     & \multicolumn{1}{c|}{69}    & 0.07 & \multicolumn{1}{c|}{11} & \multicolumn{1}{c|}{70.32} & 4.82  \\ \hline
20 & \multicolumn{1}{c|}{7}     & \multicolumn{1}{c|}{69}    & 0.0  & \multicolumn{1}{c|}{11} & \multicolumn{1}{c|}{70.89} & 3.02  \\ \hline
22 & \multicolumn{1}{c|}{7}     & \multicolumn{1}{c|}{69}    & 0.0  & \multicolumn{1}{c|}{11} & \multicolumn{1}{c|}{71.43} & 1.32  \\ \hline
24 & \multicolumn{1}{c|}{7}     & \multicolumn{1}{c|}{69}    & 0.0  & \multicolumn{1}{c|}{11} & \multicolumn{1}{c|}{71.68} & 0.86  \\ \hline
26 & \multicolumn{1}{c|}{7}     & \multicolumn{1}{c|}{69}    & 0.0  & \multicolumn{1}{c|}{11} & \multicolumn{1}{c|}{71.56} & 0.24  \\ \hline
\end{tabular}%
\end{table}

\subsubsection{Example of Relative Entropy Approach Noise Variance Estimation Illustration}
\begin{figure}[hbt!]
 	\centering
 		\includegraphics[width=8cm, trim=2.5cm 0.5cm 2.5cm 0cm]{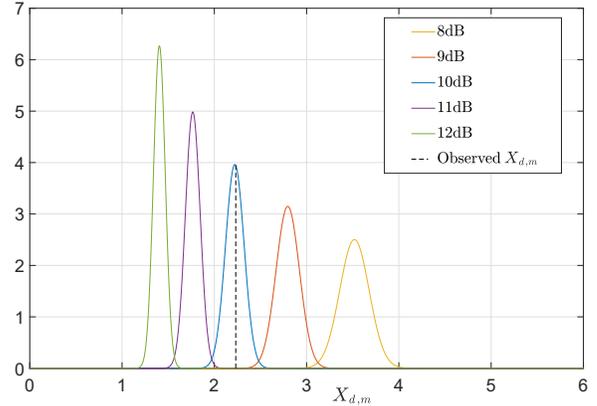}
 		\caption{Distributions of $X_{{d,m}}$ where $d=0$ and $m=48$ based on different assumptions on the value of SNR for System II (true SNR is 10dB and $X_{{0,48}} = 2.46$) }
 		\label{Fig:Xsmdistribution}
 \end{figure}

The theory of simultaneous noise variance and system modeling method is explained in Section \ref{noisevar7}. In the following we illustrate the RE based approach with an example and numbers from the IIR system.
For simplicity and without loss of generality for this illustration, lets assume that the time delay is zero.  Figure \ref{Fig:Xsmdistribution} shows the one available sample of $X_{{0,48}}$ for the IIR system and when the unknown true SNR is 10 dB. It also shows distribution of $X_{{0,48}}$ for a possible range of SNRs. Note that the calculated value of the output error $X_{{0,48}}$ for impulse length of 48 is calculated based on  (\ref{xsdmnew}) and does not require a knowledge of the SNR. 

\begin{figure}[hbt!]
 	\centering
 		\includegraphics[width=8cm,height=6.05cm,trim=1cm 0.5cm 0.7cm 0.5cm]{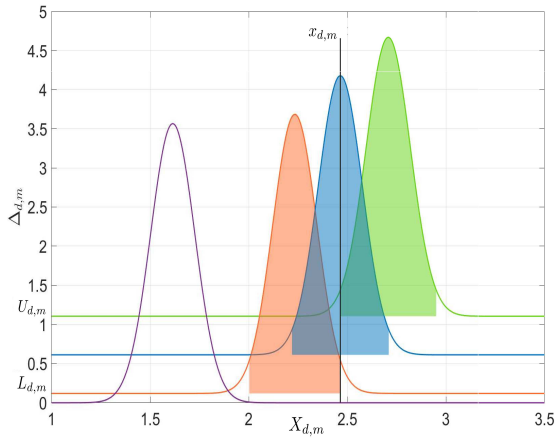}
 		\caption{Illustration of calculation of upper bound and lower bound of $\Delta_{0,48}$ based on the available $X_{{0,48}}$.} 
 		\label{Fig:Xsmbounds}
 \end{figure}

\begin{figure}[hbt!]
 	\centering
 		\includegraphics[width=8cm, trim=2.5cm 0.5cm 2.5cm 0cm]{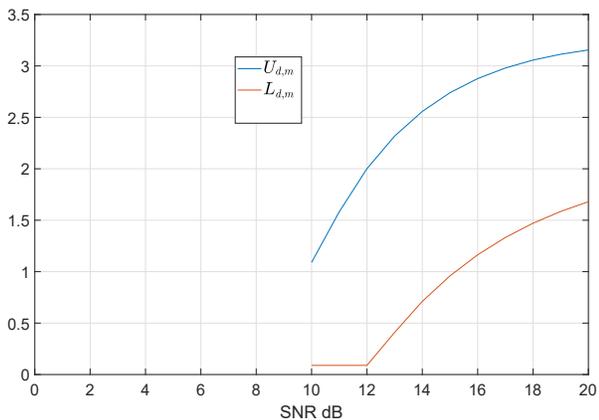}
 		\caption{Calculated $U_{{48}}$ and $L_{{48}}$ for a range of SNRs using the available $X_{{0,48}}$.}
 		\label{uplow}
 \end{figure}

\begin{figure}[hbt!]
 	\centering
 		\includegraphics[trim=90cm 1cm 90cm 2cm, scale=0.244]{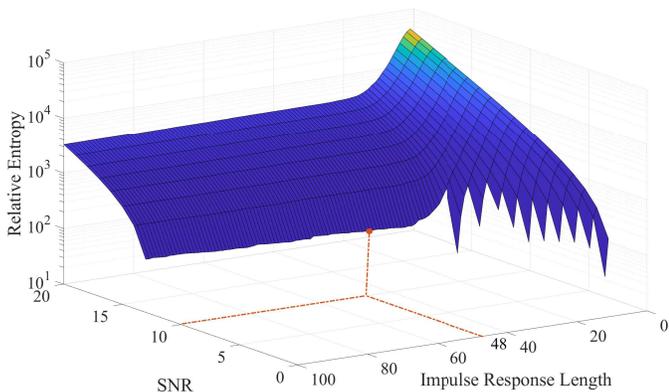}
 		\caption{Relative entropy in (\ref{eqn:dlfinal}) for a range of SNRs form 0 dB to 20 dB and for different impulse response lengths. }
 		\label{cool3d}
 \end{figure}

Figure \ref{Fig:Xsmbounds} illustrates the invalidation process of Theorem 1 and the probabilistic validation approach that finds the upper and lower bounds on  $\Delta_{0,48}$ in (\ref{eqn:unew}) and (\ref{eqn:lnew}). In this figure the distribution of $X_{{0,48}}$ is shown for a range of possible $\Delta_{0,48}$ for when the SNR is 10dB. The three colored regions under three of these distributions show the valid acceptable region for that distribution which is around its mean with the chosen validation probability of 0.999 ($\alpha=4$ in (\ref{eqn:cheby2}).  Using the one available and calculated $X_{{0,48}} = 2.46$ from the observed output, we can chose the range of $\Delta_{0,48}$ in this figure. The chosen $\Delta_{0,48}$ are the ones that include the calculated $X_{{0,48}} = 2.46$ in their validation region (the solid color). The figure shows the two extreme cases of such distributions for which the calculated  $X_{{0,48}}$ is aligned with the upper and lower bounds of these regions, i,e, the orange distribution is the smallest possible distribution of the $\Delta_{0,48}$ that can include this value and the green distribution is the largest possible one that can include this value. Therefore, the shown values of $\Delta_{0,48}$ of the boundaries of these two distributions are the desired lower and upper bounds of $\Delta_{0,48}$ which are $U_{{48}} = 1.105$ and $L_{{48}} = 0.119$.
The above procedure is repeated for a range of SNRs and consequently a range of upper and lower bounds are gathered as a function of SNR. Figure \ref{uplow} shows the result of this procedure for $m=48$ and for a range of SNR. Note that as the figure shows for this value of $m$, and by using Theorem 1, no lower and upper bounds can be calculate for SNRs less than 10db as the negative or complex values for either upper or lower bound are discarded. 
The above procedure is then repeated for a possible range of $m$ and therefore RE is calculated based on these estimated upperbounds. Figure \ref{cool3d} shows a typical behavior of the calculated relative entropy in (\ref{eqn:dlfinal}) for a range of SNRs and a range of impulse response lengths $m$. The optimum SNR and the optimum $m$ are chosen simultaneously based on relative entropy minimization. As the figure shows, the optimum SNR in this example is 10dB which is the true unknown SNR and the optimum $m$ in this example is 48. Note that in the presence of a time delay this procedure is generalized by adding a range of possible time delays as well for a simultaneous minimization of the calculated RE. 

\begin{figure}[hbt!]
 	\centering
 		\includegraphics[width=8cm, trim=2.5cm 0.5cm 2.5cm 0cm]{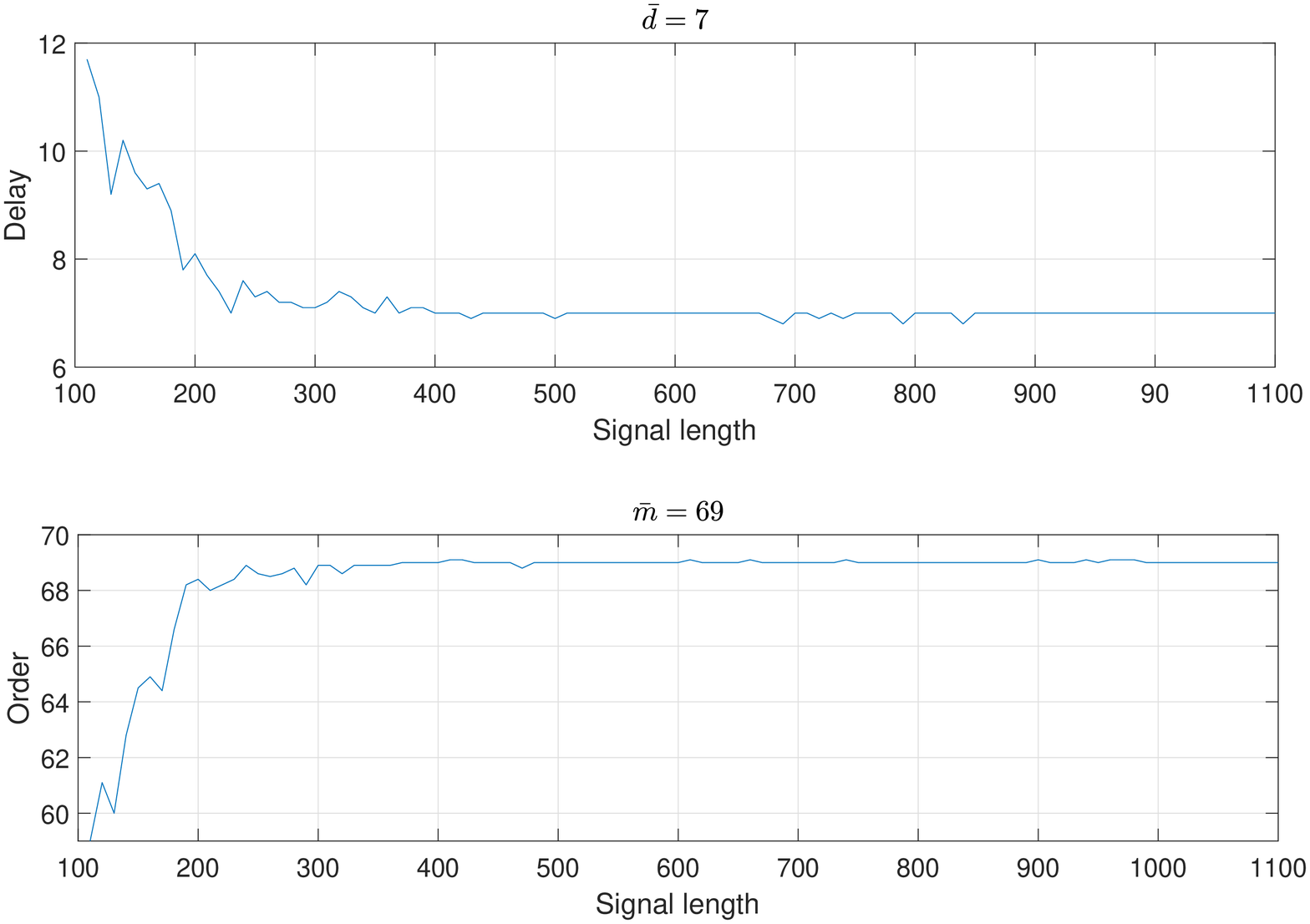}
 		\caption{Estimated online time delay and impulse response length using the online RE based method for System I, SNR=15dB} 
 		\label{Fig:stop2}
\end{figure}

\subsection{Online Modeling and Effective Role of the Stopping Criterion}
\begin{figure}[hbt!]
 	\centering
 		\includegraphics[width=8cm,trim=2.5cm 0.5cm 2.5cm 0cm]{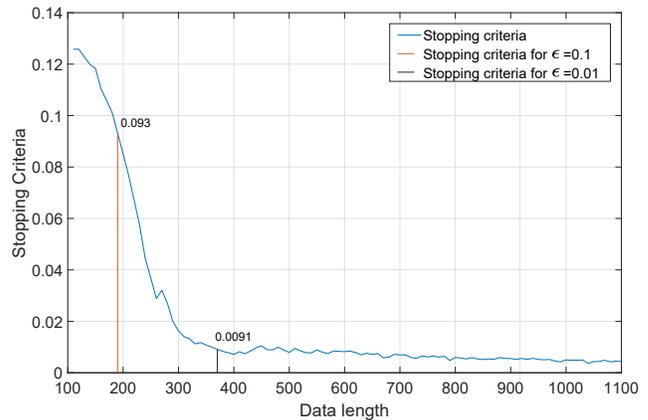}
 		\caption{Stopping criterion in online modeling of System I as the data length grows } 
 		\label{Fig:stop}
\end{figure}

Figure \ref{Fig:stop2} shows the result of the online time delay and impulse response length estimation for System I when the SNR is 15 dB, averaged over 300 trials. As the figure shows after about 400 data samples (N = 400), the online modeling converges to the true values of delay and impulse response length, which are 7 and 69. It is important to mention that in this scenario the method can be powered by the stopping criterion in practical applications, as Figure \ref{Fig:stop} illustrates.  As indicated in the figures, the stopping criterion in (\ref{eqn:Stopping}) with value $\epsilon = 0.1 $ occurs at $N = 190$. This stopping criterion is for when the desired acceptable SNR is set to $10dB$ in (\ref{eqn:epsilonSNR}). On the other hand, if the desired acceptable SNR by the user is set higher to the value of $20dB$, then the stopping criterion $\epsilon$ is $0.01$ and the algorithm automatically halts at $N=370$. In this scenario, SNR of $20dB$ waits longer and chooses the true time delay and impulse response length shown in Figure \ref{Fig:stop2}.


\section{Conclusion} \label{sec:conclusion}
A method of impulse response estimation for LTI systems, based on the theory of relative entropy (RE), is proposed. In this RE based approach time delay, impulse response length, and optimally denoised coefficients of the LTI system are estimated simultaneously. Minimizing the estimate of the relative entropy between the estimated models and true model can also provide the noise variance estimate simultaneously for when the noise variance is also unavailable. Furthermore, the extension of the proposed method for online impulse response estimations has been shown to lower the computational complexity of the estimation process. The proposed practical and efficient stopping criterion for this online LTI impulse response estimation enables the method to be used in a wide range of applications, including slowly varying LTI systems. Comparison of the proposed method with the existing time delay and order selection approaches illustrates superiority and precision of the method for a wide range of SNRs for finite and infinite impulse responses.

\appendices
{\section{Proof of Lemma 1}
Considering (\ref{17}) and (\ref{yhat}) which are elements of the reconstruction error in (\ref{zsm}), the reconstruction error is
\begin{eqnarray}
z_{{d,m}} = \frac{1}{N}||G_{{d,m}}F_{d,m}+H_{{d,m}}w^N||_2^2
\label{eqn:tot2}
\end{eqnarray}

where $w^N$ is the additive noise vector in (\ref{sys1}) and

\begin{eqnarray}
{F_{d,m}} = \left[ {\begin{array}{*{20}{c}}
{{A_{{{0,d}}}}}&{{A_{{{m,M}}}}}
\end{array}} \right]\left[ {\begin{array}{*{20}{c}}
{{\Delta _{{{0,d}}}}}\\
{{\Delta _{{{m,M}}}}}
\end{array}} \right] \label{f}
\end{eqnarray}
where $\Delta_{a,b}$ is defined in (\ref{delta}) and the matrices
\begin{eqnarray}
G_{{d,m}}=I-A_{{d,m}}(A_{{d,m}}^TA_{{d,m}})^{-1}A_{{d,m}}^T\label{eqn:gsm}  \\
H_{{d,m}}=A_{{d,m}}(A_{{d,m}}^TA_{{d,m}})^{-1}A_{{d,m}}^T \;\;\;\;\;\; \label{eqn:csm}
\end{eqnarray}
are both projection matrices. Since these projection matrices are orthogonal, the inner product of $G_{{d,m}}F_{d,m}$ and $H_{{d,m}}w^N$ is zero, and (\ref{eqn:tot2}) is
\begin{eqnarray}
z_{{d,m}} =  \underbrace{\frac{1}{N}||G_{{d,m}}F_{d,m}||_2^2}_{\Delta_{d,m}} + \frac{1}{N}||H_{{d,m}}w^N||_2^2
\label{eqn:totnew2}
\end{eqnarray}

The first term in (\ref{eqn:totnew2}) is a function of $\Delta_{d,m}$ and depends on the unmodeled parameters, which are constants.  On the other hand, the second term $\frac{1}{N}||H_{{d,m}}w^N||_2^2$ is the sum of $m-d$ zero-mean Gaussian random variables and is a non-central chi-squared random variable. Consequently, the reconstruction error $z_{{d,m}}$ is a sample of a chi-squared random variable $Z_{{d,m}}$ of order $m-d$.
Therefore, the expected value and variance of $Z_{{d,m}}$ are ( \cite{Beheshti2, beheshti2018number}):
\begin{eqnarray}
{\rm E}(Z_{{d,m}}) =
\frac{m-d}{N}\sigma_w^2+\frac{1}{N}||G_{{d,m}}F_{d,m}||_2^2 \label{eqn:aroos2}
\\ {\rm var}(Z_{{d,m}}) = \frac{2(m-d)}{N^2}(\sigma_w^2)^2 \label{eqn:doomad20} \;\;\;\;\;\;\;\;\;\;\;\;\;\;\;\;\;\;\;\;
\end{eqnarray}
Taking into account the structure of the output error in (\ref{xsdmnew}), it is also the sum of $N-(m-d)$ squares of Gaussian random variables and it can be shown that $x_{{d,m}}$ is also a sample of Chi-square random variable of order $N-(m-d)$
with the following expectation and variance \cite{Beheshti2}:
\begin{eqnarray}
E(X_{{d,m}})=(1-\frac{m-d}{N})\sigma_w^2+ \nonumber \;\;\;\;\;\;\;\;\;\;\;\;\;\;\;\;\;\;\;\;\;\;\;\;\;\;\;\; \\
\frac{1}{N}||G_{{d,m}}F_{d,m}||_2^2 \label{eqn:y22} \\
{\rm var}(X_{{d,m}})=\frac{2}{N}(1-\frac{m-d}{N})(\sigma_w^2)^2 + \;\;\;\;\;\;\;\;\;\;\;\;\;\;\;\;\;\;\;  \nonumber \\
\frac{4\sigma_w^2}{N^2}||G_{{d,m}}F_{d,m}||^2_2\label{eqn:mine}
\end{eqnarray}

\section{Proof of Theorem 1}
\label{section:thm}
Considering the Chebyshev’s Inequality  \cite{Achieser}, we have
\begin{eqnarray}
Pr\left \{ |X_{{d,m}}-E(X_{{d,m}})| \leq \nonumber
\alpha \sqrt{{\rm var}(X_{{d,m}})}    \right \} > \nonumber \\ 1 - \frac{1}{\alpha^2} \label{eqn:cheby11}
\end{eqnarray}
 where the expectation and variance of $X_{{d,m}}$ are calculated in (\ref{eqn:y22}) and (\ref{eqn:mine})  and $\alpha$ denotes the validation parameter.  The output error can be calculated by using the observed data, and thus one sample of this random variable is available. Using these values of mean and variance,  the Chebyshev’s Inequality (\ref{eqn:cheby11})  provides  probabilistic bounds on $\Delta_{d,m}=\frac{1}{N}||G_{{d,m}}F_{d,m}||_2^2$. The resulting inequality to find the upper bound are
\begin{eqnarray}
E(X_{{d,m}})-\alpha \sqrt{var(X_{{d,m}})} \leq x_{{d,m}} \label{eqn:cheby}
\end{eqnarray}
The following variables are defined for simplicity.
\begin{eqnarray}
c_{d,m} = (1-\frac{m-d}{N})\sigma_{w}^2 \;\;\;\;\;\;\; \\
\upsilon_{d,m} = 2(1-\frac{m-d}{N})(\frac{\sigma_{w}^4}{N})
\end{eqnarray}
where $\sigma_w$ denotes the standard deviation of the noise. Note that if the validation parameter $\alpha$ is chosen such that $(x_{{d,m}} - m_w) \leq -\alpha \sqrt{\upsilon_{d,m}} $ then no $\Delta_{d,m}$ term can satisfy the inequality in (\ref{eqn:cheby}). Therefore, the value of $\alpha$ must be chosen large enough, such that
\begin{eqnarray}
\alpha > \frac{N}{\sqrt{2(N-(m-d))}}\left (\frac{N-(m-d)}{N}-\frac{x_{{d,m}}}{\sigma^2_{w}}\right )
\end{eqnarray}
When solving $(x_{{d,m}} - m_w) > -\alpha \sqrt{\upsilon_{d,m}} $, the upper bound of $\Delta_{d,m}$ is the largest root of the following.
\begin{eqnarray}
\left ( \Delta_{d,m}-(x_{{d,m}} - c_{d,m}) \right ) ^2 = 
\alpha^2 \left ( \upsilon_{d,m} + \frac{4m_{w}\Delta_{d,m}}{N-(m-d)} \right )\label{sabr}
\end{eqnarray} 
which is
 \begin{eqnarray}
U_{{d,m}}=x_{{d,m}}-c_{d,m}+\frac{2\alpha^2\sigma_w^2}{N}+ \kappa_{{d,m}}(\alpha)
\label{eqn:unew12}
\end{eqnarray} 
where $\kappa_{{d,m}}(\alpha)$ is defined as
\begin{eqnarray}
\kappa_{{d,m}}(\alpha)&=&2\alpha
\frac{\sigma_w}{\sqrt{N}}\sqrt{\frac{\alpha^2\sigma_w^2}{N}+
x_{{d,m}}-\frac{1}{2}c_{d,m}}
\label{eqn:sadafam}
\end{eqnarray}

To calculate the lower bound for $\Delta_{d,m}$, the following inequality is the result of the Chebyshev inequality.
\begin{eqnarray}
 x_{{d,m}} \leq E(X_{{d,m}})+\alpha \sqrt{var(X_{{d,m}})}  \label{eqn:cheby2}
\end{eqnarray}
which is
\begin{eqnarray}
L_{{d,m}}=x_{{d,m}}-c_{d,m}+\frac{2\alpha^2\sigma_w^2}{N}- \kappa_{{d,m}}(\alpha)
\label{eqn:lnewl2}
\end{eqnarray} 

To find the upper and lower bounds of the reconstruction error, the Chebyshev inequality \cite{Achieser} is implemented:
\begin{eqnarray}
Pr\left \{ |Z_{{d,m}}-E(Z_{{d,m}})| \leq \nonumber
\beta \sqrt{{\rm var}(Z_{{d,m}})}    \right \} > \nonumber \\ 
1 - \frac{1}{\beta^2}
\end{eqnarray}
where $\beta$ denotes the confidence parameter. Taking into account the expectation and variance of the random variable $Z_{{d,m}}$ in (\ref{eqn:aroos2}) and (\ref{eqn:doomad20}) and the lower bound and upper bound calculated for $\Delta_{d,m}$ based on the observed output error in (\ref{eqn:unew12}) and (\ref{eqn:lnewl2}), the upper bound and lower bound for the reconstruction error can be calculated as
\begin{eqnarray}
\overline{z_{{d,m}}}= 2 \frac{m-d}{N}\sigma_w^2+ 
\Delta_{d,m} +\beta {\rm var}(Z_{{d,m}}) \label{upz} \;\;\;\;\;\;\;\;\;\;\;\;\ \\
\underline{z_{{{d,m}}}}=\max\{0,2 \frac{m-d}{N}\sigma_w^2+ 
\Delta_{d,m}-\beta {\rm var}(Z_{{d,m}}) \} \label{downz}
\end{eqnarray}
Using the upper bound of $\Delta_{d,m}$  in (\ref{eqn:unew12}) in the upper bound of the reconstruction error in (\ref{upz}) and the lower bound in (\ref{eqn:lnewl2}) in the lower bound of the reconstruction error in (\ref{downz}) provides the probabilistic worst-case bounds for the reconstruction error.

Note that if the order of the Chi-squared random variable is large enough (usually more than 10 is enough), it can be estimated with a Gaussian distribution. Therefore, in the case that $m-d$ is large enough, the Chi-square distribution of $Z_{{d,m}}$ and $X_{{d,m}}$ can be estimated with the Gaussian distribution. As a result, the Chebyshev inequality becomes an equality through the law of large numbers, and considering $Q(\alpha)=\int_{-\alpha}^{\alpha}\frac{1}{\sqrt{2\pi}}e^{-x^2/2}dx$ for the Gaussian equality, we have \cite{Beheshti2, beheshti2018number}
\begin{eqnarray}
{\rm Pr} \{ |X_{{d,m}}-E(X_{{d,m}})| \leq \alpha \sqrt{{\rm var} (X_{{d,m}})} \}= \nonumber \\
Q(\alpha).
\label{eqn:sadaf22}
\end{eqnarray}
\begin{eqnarray}
{\rm Pr} \{ |Z_{{d,m}}-E(Z_{{d,m}})| \leq \beta \sqrt{{\rm var} (Z_{{d,m}})} \}= \nonumber \\
Q(\beta).
\label{eqn:sadaf22}
\end{eqnarray}

\section{Proof of lemma 3}
\label{section:lem3p}
In (\ref{eqn:ANPlus}) $A^{N+1}$ is divided as follows:

\begin{eqnarray}
A_{{d,m}}^{N+1}=\left [
\begin{array}{c}
A_{{d,m}}^N\\
(B_{d,m}^N)^T 
\end{array}
\right ]
\end{eqnarray}
where 
\begin{multline}
(B_{d,m}^N)^T = \\ \left[{\begin{array}{*{20}{c}}
{{u((N + 1) - (d + 1))}}{\begin{array}{*{20}{c}}
 \cdots {{u((N + 1) - m)}}
\end{array}}
\end{array}}\right]
 \end{multline}
 
 This will update the parameter estimate in (\ref{ircestimateN}) as follows:
\begin{eqnarray}
 {\hat\theta_{{d,m}}^{N+1}} 
 =(K^{N+1}_{{d,m}})^{-1}({K^N_{S{d,m}}+ B_{d,m} B_{d,m}^{T}})^T y^{N+1}\\
 ={(K^{N+1}_{{d,m}})}^{-1}(({A_{{d,m}}^N)^T}y^N+B_{d,m}y(N+1)) 
 \label{irN}
 \end{eqnarray}
where
\begin{eqnarray}
K^N_{{d,m}} = ({A_{{d,m}}^N})^TA_{{d,m}}^N
 \end{eqnarray}
and
 \begin{eqnarray}
 K^{N+1}_{{d,m}}=K^N_{{d,m}}+ B_{d,m} B_{d,m}^{T}.
 \end{eqnarray}
The inverse of $K^{N+1}_{d, m}$ is calculated as:
 \begin{eqnarray}
(K^{N+1}_{{d,m}})^{ - 1} = {({K^N_{{d,m}}} + B_{d,m}^N{(B_{d,m}^N)^T})^{ - 1}}  \;\;\;\;\;\;\;\; \\
   = (K^N_{{d,m}})^{ - 1} - 
   \frac{1}{1 + {\rm tr}(\gamma_{d,m}^N)}C_{{d,m}}^N(C_{{d,m}}^N)^T) \label{invN}
 \end{eqnarray}
where 
\begin{align}
\label{eqn:appCL}
C_{{d,m}}^N &=(K^N_{{d,m}})^{ - 1}B_{d,m}^N\\
\gamma_{d,m}^N &=(K^N_{{d,m}})^{ - 1}B_{d,m}^N(B_{d,m}^N)^T
 \end{align}
 and tr$(a)$ is trace of matrix $a$.  Note that while $C_{{d,m}}^N$ is a vector of length $N$, $\gamma_{d,m}^N$ is a $N\times N$ matrix . In addition, it is proven in \cite{Miller} that the denominator $1 + {\rm tr}(\gamma_{d,m}^N)$ never becomes zero, and therefore this value can always be updated. 
 
The inverse update procedure in (\ref{invN}) rewrites (\ref{irN}) as follows:
 \begin{eqnarray}
 \label{eqn:appLambda}
 {\hat\theta_{{d,m}}^{N+1}}=(K^N_{{d,m}})^{ - 1} - 
   \frac{1}{1 + {\rm tr}(\gamma_{d,m}^N)}C_{{d,m}}^N(C_{{d,m}}^N)^T \times \nonumber \\
 (({A_{{d,m}}^N})^Ty^N+B_{d,m}y(N+1)) 
\\
 =(I-\frac{1}{1 + {\rm tr}(\gamma_{d,m}^N)}\gamma_{{d,m}}^N)\hat \theta _{{{d,m}}}^{N}+ \nonumber\\
 (I+\gamma_{d,m}^N)C_{{d,m}}^Ny(N+1)
 \end{eqnarray}
}

Since $\gamma_{d,m}^N$ is  a $N \times N$ matrix (calculated in (\ref{eqn:appCL})) and $C_{{d,m}}^N$ is a vector of length $N$ (calculated in (\ref{eqn:appCL})), the complexity order of the recursive calculation of $\hat \theta_{d,m}^{N+1}$ with respect to (\ref{eqn:appLambda}) is of order $O(N^2)$. 


%
\bibliography{mybibfile}

\vspace{11pt}


\vfill

\end{document}